\documentclass{edbk}
\usepackage{edbkps_unix}
\usepackage{epsfig}
\let\footnote\savefootnote
\let\footnoterule\savefootnoterule 
\normallatexbib

\begin{document}

\articletitle{Stability of d-wave superconductivity in the $\lowercase{t}{-}J$ model}

\author{Federico Becca, Luca Capriotti, and Sandro Sorella}
\affil{Istituto Nazionale per la Fisica della Materia (INFM) and
International School for Advanced Studies (SISSA),
Via Beirut 4, I-34013 Trieste, Italy }

\begin{abstract}
We use a recently developed technique, which allows to perform few Lanczos steps
on a given wavefunction even for large system sizes,
to investigate the $t{-}J$ model in the physical parameter region
and to check the stability of the BCS d-wave variational
wavefunction \cite{gros}.
Our statistical Lanczos algorithm, which extends and improves the
one Lanczos step proposed in Ref.~\cite{heeb}, has been extensively
tested on the small $L=18$ sites cluster where many  Lanczos iterations can be performed exactly.
In this case, at doping $\delta\sim 10\%$ the BCS wavefunction represents
a very good initial state to achieve extremely accurate energies and 
correlation functions with few Lanczos iterations.
For large sizes ($L\le 98$) the behavior is  similar: the
low-energy d-wave order parameter $P_d$ is weakly affected by a couple 
of Lanczos iterations in the low doping $\delta\sim 10\%$ region,
whereas the energy is considerably lowered.
As a further test of our calculation we have computed the variance of
the Hamiltonian  
$\Delta E_p=(\langle \hat{H}^2 \rangle-\langle \hat{H} \rangle^2)/L^2$  
on the BCS wavefunction with $p=0,1,2$ Lanczos steps. 
For large $p$, when the Lanczos algorithm converges to the exact
ground state, the variance vanishes exponentially with increasing $p$.
The remarkable reduction of the variance, observed for $p=1,2$ Lanczos steps
even for the largest lattice size considered, suggests
a smooth and rapid convergence to the exact ground state.
These results support the existence of off-diagonal long-range d-wave 
superconducting order in the two-dimensional $t{-}J$ model.
\end{abstract}

\section{Introduction}

One of the most important question raised after the discovery of
high-Tc superconductivity is whether a simple model of strongly correlated
electrons can capture the low-energy physics of real materials.
In particular it is still a very much debated issue 
whether a purely repulsive electronic interaction can give rise to
a d-wave superconducting ground state by doping an antiferromagnetic Mott insulator
with a small amount of holes.
The simplest model that has been proposed immediately after the 
discovery of high-Tc superconductors is the $t{-}J$ model \cite{anderson,zhang}
\begin{equation}\label{tj}
\hat{H}= J \sum_{\langle i,j \rangle} \left ( \hat{{\bf S}}_i \cdot \hat{{\bf S}}_j -
\frac{1}{4} \hat{n}_i \hat{n}_j \right )
-t \sum_{\langle i,j \rangle, \sigma}
{\tilde c}^{\dag}_{i,\sigma} {\tilde c}_{j,\sigma},
\end{equation}
where ${\tilde c}^{\dag}_{i,\sigma}=\hat{c}^{\dag}_{i,\sigma}
\left ( 1- \hat{n}_{i,\bar \sigma} \right )$,
$\hat{n}_i= \sum_{\sigma} \hat{n}_{i,\sigma}$ is the electron
density on site $i$,
$\hat{{\bf S}}_i=\sum_{\sigma,\sigma^{\prime}}
{\tilde c}^{\dag}_{i,\sigma} {\bf \tau}_{\sigma,\sigma^{\prime}}
{\tilde c}_{i,\sigma^{\prime}}$ is the spin operator and
${\bf \tau}_{\sigma,\sigma^{\prime}}$ are Pauli matrices. In the following
we put $t=1$.

After many years of intense numerical and theoretical efforts  
there is no general consensus on the properties  
of this simple Hamiltonian and of the related Hubbard model.
From the numerical point of view, the density matrix renormalization group
(DMRG) \cite{dmrg} predicts that a charge density wave instability \cite{white1},
nowadays called {\em striped-phase} for its one dimensional character, is strongly 
competing with pairing and superconductivity \cite{white2}. 
Within DMRG it appears therefore difficult to explain Copper-Oxide 
superconductors with a simple one-band model, especially because, 
in order to be consistent with photoemission
experiments \cite{shen}, a negative next-nearest-neighbor hopping 
amplitude $t^\prime$ has to be included in the model.
In fact, the negative $t^\prime$ suppresses even further superconductivity, 
so that the $t{-}J$ model becomes unrealistic
to describe the low-energy physics of high-Tc superconductors. 
However, the DMRG results, though quite accurate, 
are not exact in two dimensions.
Moreover, for technical reasons it is possible to consider only particular
boundary conditions (open in one direction and periodic in the other),
which certainly make the DMRG calculation still far to be representative of
the thermodynamic limit. 

Quantum Monte Carlo (QMC) is an appealing alternative numerical approach.
This numerical method is still severely limited for two-dimensional fermionic systems  
by the well-known {\em sign problem} and is 
consequently biased by the initial guess of the ground state 
used to control this numerical instability \cite{fn,sorella,caprio}.
However this technique has the important advantage to work very well
with periodic boundary conditions  since translation invariance
can be explicitly used to improve the efficiency of QMC algorithms.
In particular, an approximate ground state can be obtained starting from
a translation invariant wavefunction $|\psi_G\rangle$
by applying exactly few powers of the 
Hamiltonian $(-\hat{H})^p$ \cite{chen} or many approximate ones 
(by using for instance the {\em Fixed-Node} FN approximation \cite{fn}).
However, within various QMC schemes the situation is still controversial. 
First Heeb \cite{heeb} and Khono \cite{khono} have found  d-wave 
superconductivity in a reasonable 
parameter range. Later Shih and {\em co-workers} \cite{tklee}, using 
a very similar method, have excluded drastically this possibility.
The latter results were obtained within the monotonic-behavior assumption of 
the off-diagonal superconducting order parameter as a function 
of the number $p$ of Hamiltonian powers applied to the initial wavefunction.
This assumption, although reasonable, is highly questionable.
By contrast, in a recent QMC work \cite{calandra},
a clear tendency to d-wave superconductivity in the $t{-}J$ model was found. 
Moreover, very recently, an almost realistic phase diagram with a 
corresponding high-Tc d-wave superconducting transition has been obtained
for the Hubbard model within the Dynamical Mean Field Approximation \cite{lich,pruschke}.
Furthermore, in the contest of the Hubbard model, also a weak-coupling 
renormalization group approach \cite{rice,metzner} gives rise to a d-wave
order parameter in a large region of the phase diagram.
The latter results strongly support the relevance of 
a single band model for  the explanation of  high-Tc superconductivity.
 
\section{Numerical method}

In this work we make a further attempt to clarify the controversial 
numerical findings on the issue of d-wave superconductivity
in the $t{-}J$ model, using
the statistical {\em few Lanczos-step technique} (FLST), 
efficiently implemented by means of the 
stochastic reconfiguration (SR) \cite{sorella,caprio}.
Within the latter scheme all kind of correlation functions 
can be computed efficiently without any {\em mixed average} 
\cite{caprio} approximation:
an enormous advantage compared to the FN or to the original  
SR technique \cite{sorella,caprio}. In these cases  
this bias can be removed at the price of adding a small field coupled to the 
desired correlation function \cite{calandra}. However, it turns out
that it is extremely difficult and 
computationally demanding to work in the small field limit when unbiased 
correlation functions can be obtained.
FLST is instead a very good compromise that solves efficiently 
this numerical problem of QMC methods. 

The wavefunctions that we are able to sample statistically read:
\begin{equation} \label{psig}
|\psi_{p}\rangle =\Big(1+\sum\limits_{k=1}^p  \alpha_k \hat{H}^p \Big) 
|\psi_G\rangle
\end{equation}
with parameters $\{ \alpha_k\}$ for $k=1,\cdots, p$ minimizing the energy 
expectation value 
$\langle\psi_p|\hat{H}|\psi_p\rangle / \langle\psi_p|\psi_p\rangle$. 
For any $p$ it is simple to show that the wavefunction (\ref{psig}) 
corresponds exactly to apply 
$p$ Lanczos step iterations to the initial wavefunction $|\psi_G\rangle$.
This wavefunction is sampled statistically with the SR technique, 
by using in the reconfiguration scheme 
the first $p$ powers of the Hamiltonian. 
Unlike the previous method \cite{sorella,caprio} 
the reference wavefunction $|\psi^f\rangle$ 
is not evolved during  the Markov iteration. $|\psi^f\rangle$ is instead
kept statistically equal to the initial wavefunction 
$|\psi_G\rangle$ using 
the variational scheme proposed by Hellberg and Manousakis \cite{manusakis},
which highly reduce the statistical fluctuations related to the SR technique.
The equivalence of FLST to the standard Lanczos algorithm
will be discussed in a forthcoming paper \cite{tobepublished}. 

The initial wavefunction to which FLST will be applied 
can be written as follows \cite{gros}:
\begin{equation} \label{gros}
|\psi_G \rangle=|\psi_{p=0} \rangle = \hat{P}_0 \, \hat{P}_N \hat{J} |D\rangle.
\end{equation}
where $|D\rangle$ is a BCS wavefunction, which is an exact eigenstate of the 
following Hamiltonian:
\begin{eqnarray}
\hat{H}_{BCS}&=&\hat{H}_0+ \frac{\Delta_{BCS}}{2} 
(\hat{\Delta}^\dag +\hat{\Delta})  \label{hbcs} \\
\hat{\Delta}^\dag&=& \sum_{\langle i,j \rangle} M_{i,j} \big( 
 \tilde{c}^\dag_{i,\uparrow} \tilde{c}^{\dag}_{j,\downarrow} + 
 \tilde{c}^\dag_{j,\uparrow} \tilde{c}^{\dag}_{i,\downarrow} \label{ddag} \big) 
\end{eqnarray}
where $\hat{H}_0=\sum\limits_{k,\sigma} \epsilon_k \, 
\tilde{c}^\dag_{k,\sigma} \tilde{c}_{k,\sigma}$ 
is the free electron tight binding nearest-neighbor Hamiltonian, 
$\epsilon_k=-2(\cos k_x + \cos k_y) -\mu$, $\mu$ is the free-electron 
chemical potential and  
$\hat{\Delta}^\dag$ creates all possible nearest-neighbor singlet bonds 
with  d-wave symmetry being $M_{i,j}$ $+1$
or $-1$ if the bond $\langle i,j \rangle$ is in the $x$ or $y$ 
direction, respectively.
$\hat{P}_N$ and $\hat{P}_0$ are the projectors over the subspaces with
a fixed number $N$ of particles and no doubly occupied states.
Finally the Jastrow factor 
$\hat{J}=\exp\left( \gamma/2 \sum_{i,j} v(i-j) 
\hat{n}_i \hat{n}_{j} \right)$ couples the 
holes via the density operators $\hat{n}_i$ and 
contains another variational parameter $\gamma \sim 1$ which scales an 
exact analytic form, obtained by  approximating   
the holes  with hard-core bosons at the same density, 
and applying the spin-wave theory
to the corresponding $XY$ model \cite{franjic}.
We note here that by performing a particle-hole transformation on the 
spin down $\tilde{c}^\dag_{i,\downarrow} \to (-1)^i \tilde{c}_{i,\downarrow}$, 
the ground state of the BCS Hamiltonian is just a Slater-determinant with 
$N=L$ particles \cite{shiba}. This is the reason why this variational wavefunction
can be considered of the  generic Jastrow-Slater form, a standard 
variational wavefunction used in QMC. Using
the particle-hole transformation, it is also possible to control exactly
the spurious finite system divergences related to the nodes of the 
d-wave order parameter.
 
\section{Numerical tests}

In this section we show the accuracy of FLST applied to the BCS wavefunction 
(\ref{gros}) on a small 18-site cluster, where exact results are available.

\begin{figure}[htb]
\center{\epsfig{file=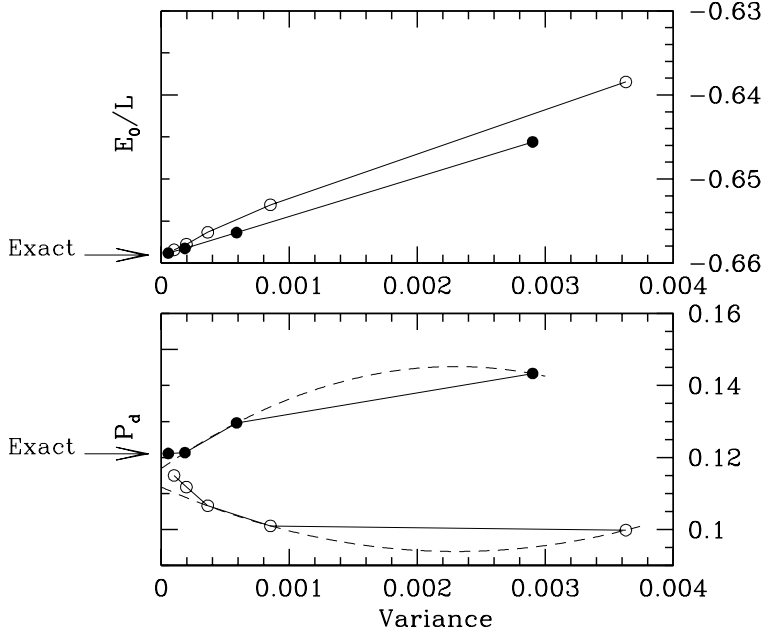,height=110mm,angle=0}}
\caption{Ground-state energy per site $E_0/L$ and d-wave order parameter $P_d$
as a function of the variance for $N=16$, $L=18$, $J=0.4$: full dots (optimal $\Delta_{BCS}$),
empty dots ($\Delta_{BCS}\to 0)$. 
Dashed lines are quadratic fits of the estimates with $p=0,1,2$.}
\label{ergpd18}
\end{figure}

Our main task is to compute the order parameter at finite system size 
\begin{equation}\label{pd} 
P_d={1\over L} \langle \psi_p^{N+2}| \hat{\Delta}^\dag |\psi_p^N\rangle,
\end{equation}
where $|\psi_p^N\rangle$ and $|\psi_p^{N+2}\rangle$ are the states with $N$
and $N+2$ particles, respectively. If $P_d$ is finite in the thermodynamic 
limit this necessarily implies off-diagonal 
long-range order in the ground state. 
Following Ref.~\cite{calandra}, it is convenient with an approximate
technique to calculate a short-range quantity like $P_d$,
instead of the more conventional long-range expectation value
$\langle \psi_p^N| \hat{\Delta} \hat{\Delta}^\dag |\psi_p^N \rangle /L^2 $. 
In Table \ref{tone} we show a comparison between FLST and the exact results for 18 and 
16 electrons at $J=0.4$.

In Table \ref{ttwo} we show $P_d$ 
as a function of the number of Lanczos step iterations for the 
18-site cluster at $J=0.4$.
In the same Table we have computed also the variance 
$\Delta E_p=(\langle\psi_p |\hat{H}^2|\psi_p\rangle  
-\langle\psi_p|\hat{H}|\psi_p\rangle ^2)/L^2$, 
the overlap squared $Z_p=|\langle\psi_p|\psi_0\rangle |^2$ of 
the FLST  wavefunction  with the true ground state 
$|\psi_0\rangle $, and the average sign of the FLST wavefunction: 
\begin{equation} \label{sign}
\langle S_p\rangle =\sum_x \langle x|\psi_0\rangle^2 
{\rm Sgn} \left( \langle x|\psi_p\rangle \langle x|\psi_0\rangle \right)~,
\end{equation}
where $|x\rangle$ denotes configurations with  definite electron positions 
and spins.
For an exact calculation, namely  $p >> 1$, both $ Z_p \to 1$ 
and $S_p\to 1$, whereas $ \Delta E_p \to 0$.
The variance thus represents a very important tool to estimate the 
`distance' from the exact ground state when the latter one is not known. 
In particular whenever $Z_p \simeq  1$ 
the energy approaches the exact result linearly with the 
variance $\Delta E_p$, allowing us to estimate the error 
in the variational energy. 

\begin{table}
\begin{tabular}{ccccccc}
$N$ & $\Delta_{BCS}$ & $p$ & $E_0/L$ (FLST) & $P_d$ (FLST) & $E_0/L$ (Exact) & $P_d$ (Exact)\\
\hline\hline
18  &       0.55     &  1  &  -0.4765(1)    &              &   -0.47668      &            \\
18  &       0.55     &  2  &  -0.4775(1)    &              &   -0.47749      &            \\
16  &       0.20     &  1  &  -0.6541(1)    &  0.1074(4)   &   -0.65420      &   0.10730  \\
16  &       0.55     &  2  &  -0.6583(1)    &  0.122(1)    &   -0.65826      &   0.12135  \\
\hline\hline
\end{tabular}
\caption{Comparison between the estimates of the ground-state energy per site
$E_0/L$ and of the d-wave order parameter $P_d$ obtained with the
exact and the statistical (FLST) application of $p$ Lanczos steps
on the variational wavefunction of Eq.~(\ref{gros}).  $L=18$, $N=16,18$ and $J=0.4$.}
\label{tone}
\end{table}

\begin{table}
\begin{tabular}{ccccccc}
$N$  &  $\Delta_{BCS}$ &  $p$ & $\langle S_p \rangle$ &  $Z_p$ & $\Delta E_p\times L^2$ & $E_0/L$ \\
\hline\hline
18   &     0.00        &   0    &   1.0000              & 0.6898 &  1.194      & -0.43833  \\
\hline
16   &     0.00        &   0    &   0.9656              & 0.8306 &  1.174      & -0.63847  \\
\hline
18   &     0.80        &   0  &   1.0000              & 0.8850 &     0.335     & -0.46639  \\
18   &     0.80        &   1  &   1.0000              & 0.9915 &     0.042     & -0.47662  \\
18   &     0.80        &   2  &   1.0000              & 0.9995 &     0.004     & -0.47752  \\
18   &     0.80        &   3  &   1.0000              & 1.0000 &     0.0003    & -0.47759  \\
18   &     0.80        &   4  &   1.0000              & 1.0000 &     0.00002   & -0.47759  \\
18   &     0.80        &$\infty$&   1.0000              & 1.0000 &     0.0       & -0.47759  \\
\hline
16   &     0.55        &   0    &   0.9891              & 0.9260 &     0.940     & -0.64559  \\
16   &     0.55        &   1    &   0.9988              & 0.9814 &     0.191     & -0.65638  \\
16   &     0.55        &   2    &   0.9999              & 0.9942 &     0.060     & -0.65826  \\
16   &     0.55        &   3    &   1.0000              & 0.9983 &     0.018     & -0.65882  \\
16   &     0.55        &   4    &   1.0000              & 0.9995 &     0.005     & -0.65898  \\
16   &     0.55        &   5    &   1.0000              & 0.9999 &     0.002     & -0.65902  \\
16   &     0.55        &   6    &   1.0000              & 0.9999 &     0.0005    & -0.65904  \\
16   &     0.55        &   7    &   1.0000              & 1.0000 &     0.0001    & -0.65904  \\
16   &     0.55        &$\infty$&   1.0000              & 1.0000 &     0.0       & -0.65904  \\
\hline\hline
\end{tabular}
\caption{Average sign $\langle S_p \rangle$, overlap squared on the exact ground state $Z_p$
and variance times the volume squared $\Delta E_p\times L^2$ obtained applying exactly
$p$ Lanczos steps on the variational wavefunction of Eq.~(\ref{gros}).
$L=18$, $N=16,18$ and $J=0.4$.}
\label{ttwo}
\end{table}

This can be achieved by plotting the  variational energies 
$E_p$ as a function of the corresponding variance $\Delta E_p$, 
and performing a very stable linear or quadratic
fit to the $\Delta E_p=0$ exact limit (see Fig.~\ref{ergpd18}). 
Similar fits can be attempted for 
correlation functions though, in this case, also a term 
$\propto \sqrt{ \Delta E_p} $ is 
expected for $\Delta E_p \to 0$. This term is however negligible for  
quantities like $P_d$ that are averaged bulk correlation functions 
in a large system size (see the Appendix).
In practice even in the small 18-site cluster 
the non-linear term turns out to be negligible (see Fig.~\ref{ergpd18}).
We believe that, being the convergence of 
the Lanczos  algorithm particularly well behaved and certainly unbiased, 
the variance extrapolation 
method is in this case particularly useful and reliable. 
However for bad initial wavefunction (e.g., randomly generated) or
very large sizes the approach to zero of the variance may behave 
rather wildly, requiring many Lanczos steps to reach the regime 
where the extrapolation is possible.

As shown in Table \ref{ttwo} the quality of the variational 
BCS wavefunction (\ref{psig}) is {\em exceptionally good}, especially
in the doped $N=16$ case. Here $Z_p$ is larger 
than $0.9$ even at the simplest $p=0$ variational level, and is 
drastically improved with really few Lanczos step iterations. 
Remarkable is also the behavior of the average sign $S_p$ which measures 
directly the accuracy of  the BCS wavefunction phases, without caring about
the amplitudes.
In the undoped case the signs of the BCS wavefunction $\langle S_0 \rangle$ 
can be proven to be exact, i.e., $\langle S_0 \rangle = 1$, having 
the BCS state the well-known Marshall
signs, i.e., the phases of the exact ground state of Heisenberg model.
For the two-hole case, the BCS nodes change in a non trivial way. 
Nevertheless,  $\langle S_0 \rangle$ remains very close to $1$ and it is much 
higher than the average sign of the 
corresponding Gutzwiller wavefunction ($\Delta_{BCS} \to 0$),  
also shown in the table for comparison.

These results suggest that there is a tendency to d-wave BCS pairing  
in the $t{-}J$ model at $\sim 10\%$ doping and $J \sim 0.4$, 
and that the BCS  wavefunction is a particularly accurate wavefunction
to describe the small and even zero doping region of the $t{-}J$ model.

\section{Larger size calculations}

\begin{figure}[htb]
\center{\epsfig{file=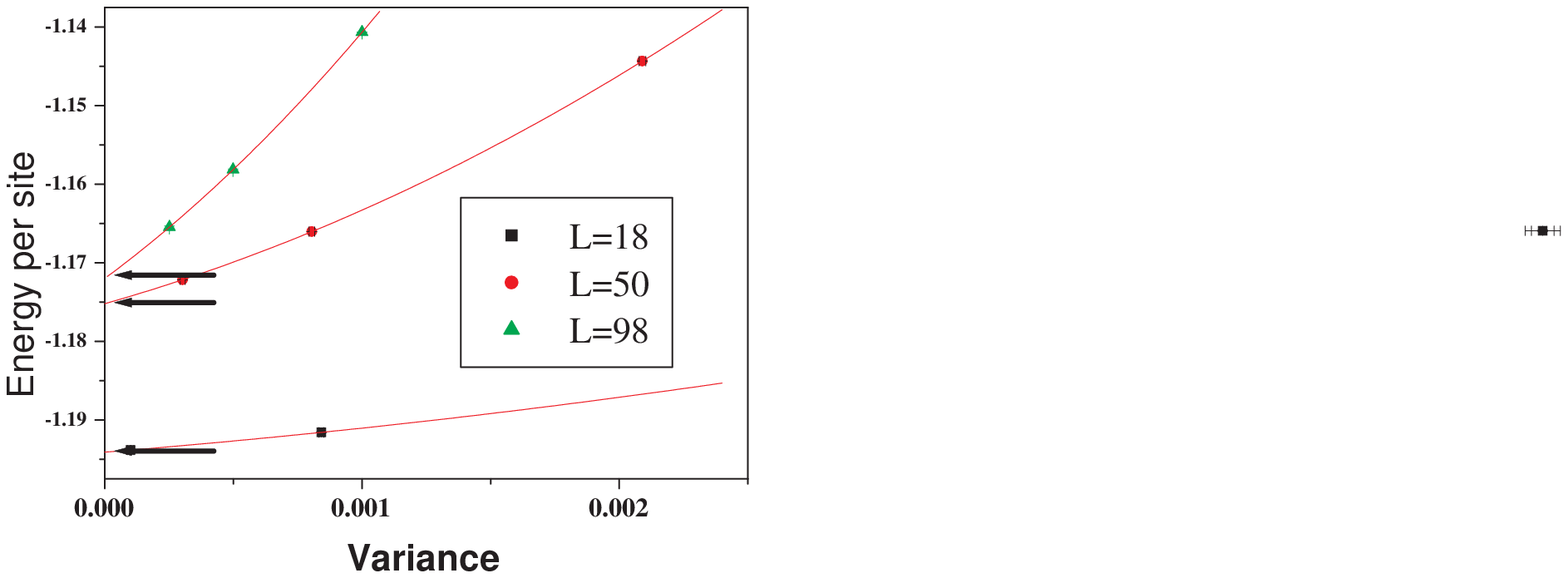,height=80mm,angle=0}}
\caption{Energy per site of the finite-size Heisenberg model. 
Comparison of exact results (indicated by arrows) and the approximate $p=0,1,2$ Lanczos step
iterations  over the projected d-wave wavefunction.
Continuous lines are quadratic fit of the data.}
\label{figcav}
\end{figure}

\begin{figure}[htb]
\center{\epsfig{file=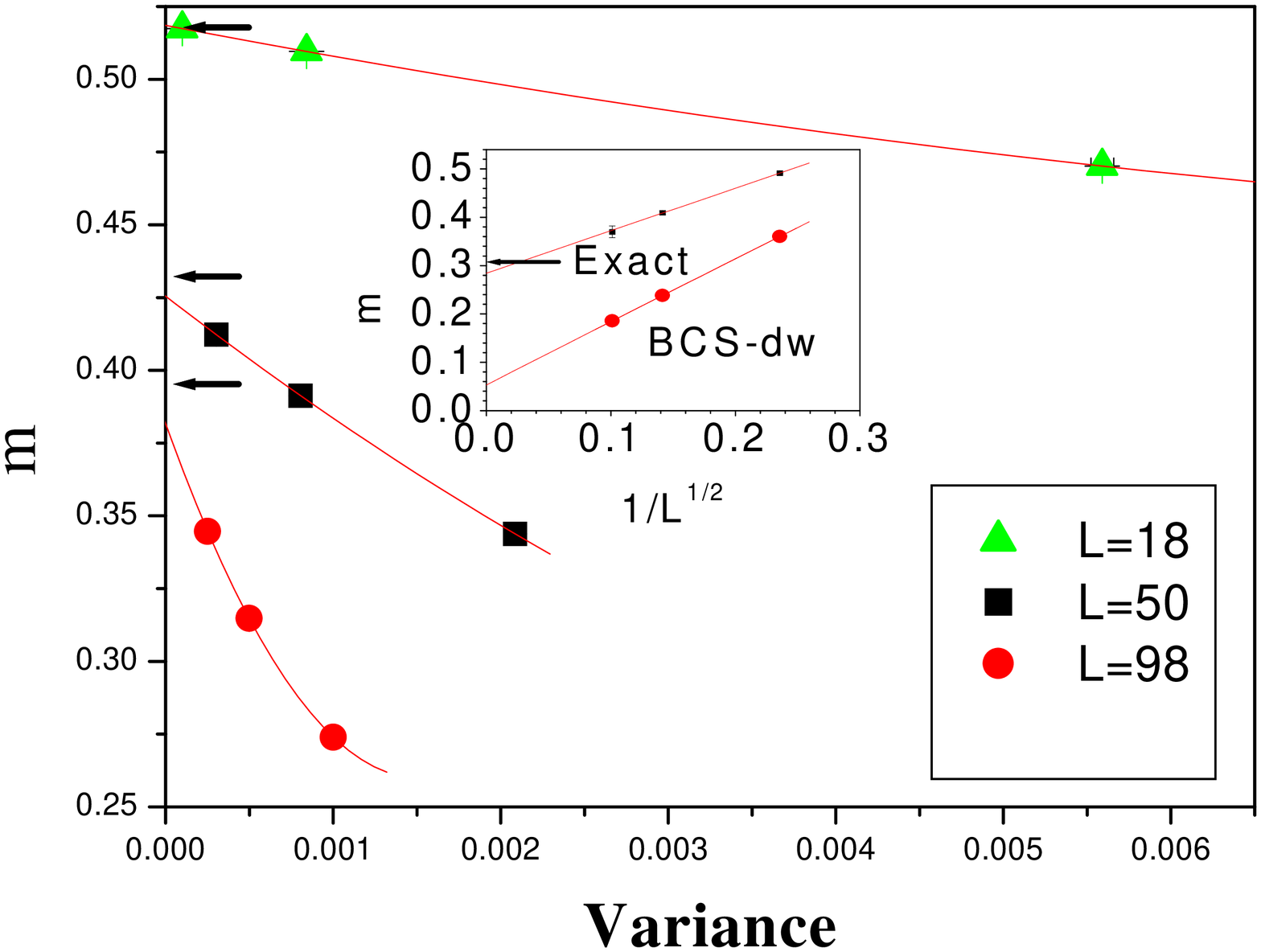,height=80mm,angle=0}}
\caption{Order parameter
$m=\protect\sqrt{S(\pi,\pi)/L}$ in the finite-size 
Heisenberg model ($S(\pi,\pi)$ being the spin isotropic
antiferromagnetic structure factor). Comparison of exact results
(indicated by arrows) and the approximate $p=0,1,2$ Lanczos step
iterations over the projected d-wave wavefunction.
Continuous lines are quadratic fit of the data.  Inset: finite-size scaling
with the variational (BCS d-wave) wavefunction and with 
the variance extrapolated one.}
\label{figcavm}
\end{figure}

Though few  Lanczos steps  may appear inadequate  for large system size,
this simple scheme is instead providing us very good variational energies  
up to  $L\sim 100$ sites in the  $t{-}J$ model, even when this 
variational energies are compared with more complicated schemes like the 
FN.
In order to show that the Lanczos scheme  remains  effective 
for larger sizes it is useful to consider first 
a relevant case where  a  numerically exact  solution is possible: 
the zero doping limit of the $t{-}J$ model, i.e., the  Heisenberg model. 
In Fig.~\ref{figcav} we plot the energy results calculated for few Lanczos 
steps as a function of the variance. The fact that the energy is approaching  the exact
result smoothly with the variance both in the 50- and 98-site clusters
(with a slightly larger curvature in the latter case) 
indicates that the projected BCS wavefunction should have 
a substantially large overlap squared $Z_p$ with the exact ground state
of the Heisenberg model even on these large system sizes.
This is also confirmed by the behavior of the square order parameter 
(Fig.~\ref{figcavm}), which is considerably improved by few Lanczos steps. 
This result is particularly important since the starting BCS 
wavefunction has not antiferromagnetic long-range order, 
whereas the two-dimensional Heisenberg model is widely believed to be antiferromagnetically 
ordered.

The above test represents also a further 
strong evidence that the ground-state wavefunction of the Heisenberg model is 
smoothly connected to a d-wave BCS superconducting wavefunction. 
This circumstance represents a very interesting   
numerical fact that clearly supports the experimental observation 
of high-Tc d-wave superconductivity coming just upon 
a small doping of a quantum antiferromagnet.

\begin{figure}[htb]
\center{\epsfig{file=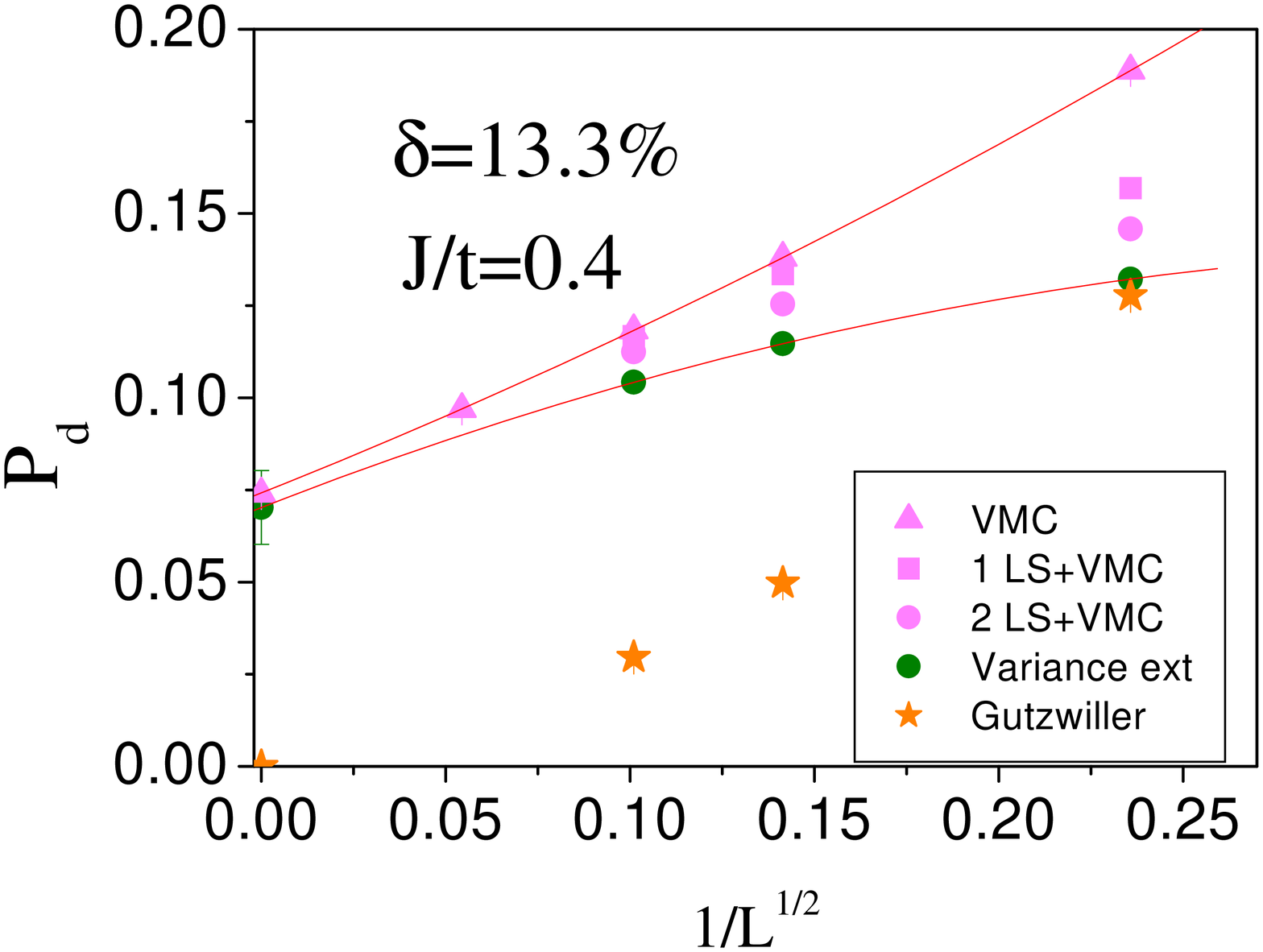,height=80mm,angle=0}}
\caption{Superconducting  d-wave order parameter $P_d$ in the
$t{-}J$ model as defined in the text in Eq.~(\ref{pd}). These results were obtained
starting by the variational wavefunction (triangles) defined in Eq.~(\ref{gros}) and
by applying to it one (squares) or two (dots) Lanczos steps.
Black dots are obtained by quadratic extrapolation to the zero variance exact limit
in order to estimate the size dependent error of the approximate variational
calculations. The stars refer to the Gutzwiller wavefunction
with $\Delta_{BCS}\to0$ and $\gamma=1$ in Eq.~(\ref{gros}).
Continuous lines are quadratic fit of the data.}
\label{figpd}
\end{figure}

\begin{figure}[htb]
\center{\epsfig{file=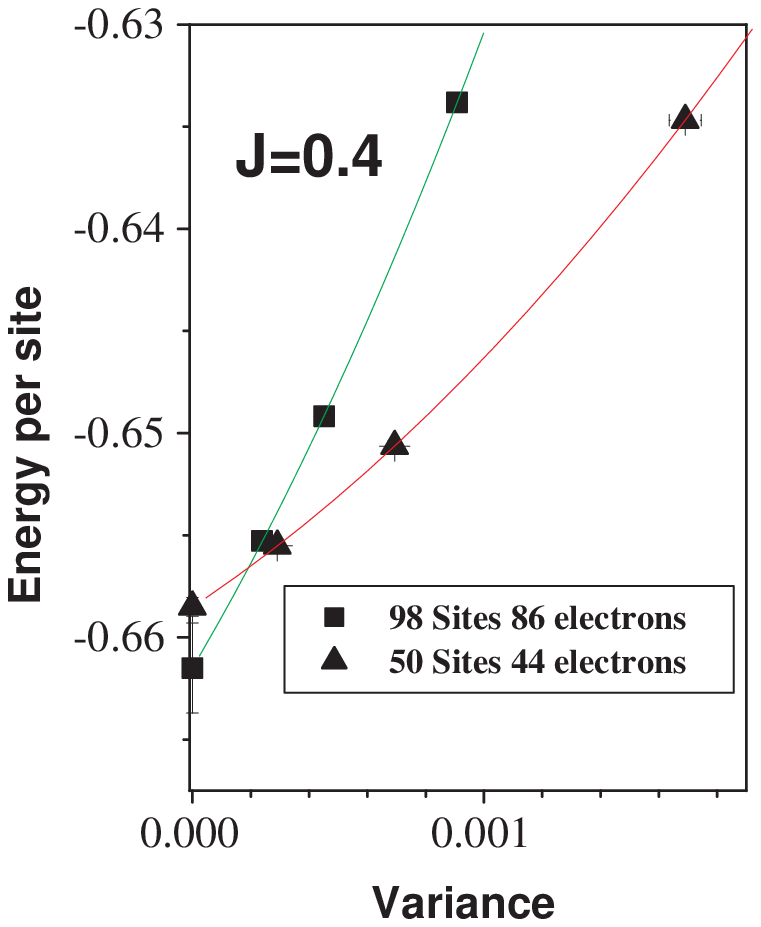,height=80mm,angle=0}}
\caption{Energy per site of the $t-J$ model for 44 electrons on 50 sites (triangles) 
and 86 electrons on 98 sites (squares), $J=0.4$, for the approximate $p=0,1,2$ Lanczos step
iterations over the projected d-wave wavefunction Eq.~(\ref{gros}). 
Continuous lines are quadratic fit of the data.}
\label{edoped}
\end{figure}

Within the Lanczos approximate states $|\psi_p\rangle$ ($p=0,1,2$) 
acting on the best variational BCS wavefunction, we have performed 
a finite-size scaling of the order parameter $P_d$ as defined in 
Eq.~(\ref{pd}), at a fixed doping $\delta=13.3\%$ (corresponding to 
$N=84$ on the largest size $L=98$).
Since $P_d$ is computed between two states 
with $N$ and $N+2$ particle numbers, we assume that 
it refers to the intermediate doping  $\delta=1-(N+1)/L$.
For smaller sizes the doping $\delta=13.3\%$ is not possible and we have
interpolated linearly $P_d$ between the two fillings closest to this doping.
The main results of this paper is then shown in the Fig.~\ref{figpd}.
Here the size scaling for $p=0,1,2$ Lanczos steps and for their zero-variance
extrapolation clearly indicates a finite $P_d$ in the thermodynamic limit. 
The stability of the BCS variational wavefunction is evident from this figure: 
$P_d$ remains much larger than the corresponding value of the Gutzwiller 
($\Delta_{BCS} \to 0$) metallic state.
Instead,  as shown in Fig.~\ref{edoped} and similarly to
the undoped case, the energy is very much lowered
by FLST, suggesting that the method remains effective even  for large sizes
and finite doping. 

\section{Conclusions}

In conclusion we have provided evidences in favor of the stability of d-wave pairing 
correlations in the $t{-}J$ model.
Our result disagrees with a recent QMC calculation \cite{tklee},
for which, however, the self-consistent assumption mentioned in the 
introduction leads to a very poor, high-energy variational state. 
Furthermore, also the DMRG results provide a qualitatively different scenario, 
namely that stripes suppress d-wave superconductivity \cite{white2}.
In order to clarify this disagreement, we have performed a simulation 
at $J=0.4$ for a cluster with $L=6\times 12$, 8 holes, and open boundary 
conditions on the long direction, as  usually done within DMRG.
Our accuracy is in this case  worse than 
with periodic boundary condition, 
especially for quantities that are not bulk 
average correlation functions (see the Appendix). 
For $P_d$ instead we have found that, even at the simplest variational level 
open  boundary conditions strongly suppress $P_d$ by more than $30\%$, much more
than the difference found in Fig.~(\ref{figpd}).
Thus we find that boundary conditions play 
a very important role and may strongly destabilize the uniform BCS d-wave 
ground state, as obtained within DMRG. 
This is consistent with experimental findings \cite{tranquada},
showing  that any small asymmetry in the CuO planes, 
acting in our point of view as a distorted boundary condition,
may enhance the tendency to stripe formation and then suppressing superconductivity. 
Our results instead support the recent
numerical works \cite{lich,pruschke,rice,metzner}, indicating
that d-wave superconductivity can be obtained in a one band model with 
repulsive electron interaction. 

\chapappendix{Variance estimate of the error on bulk correlation functions}
In this Appendix we  estimate the error on correlation functions 
assuming that the ground state $|\psi_0\rangle $ is approximated with the 
wavefunction $|\psi_p\rangle $:  
\begin{equation}\label{defeps}
|\psi_0\rangle =|\psi_p\rangle +\epsilon_p |\psi^\prime\rangle 
\end{equation}
where $\langle \psi_p|\psi_p\rangle =\langle \psi^\prime|\psi^\prime\rangle =1$, 
and $|\psi^\prime\rangle$ 
represents a normalized wavefunction orthogonal to the exact one, 
$\langle \psi_0|\psi^\prime\rangle =0$.
We restrict our analysis to thermodynamically averaged correlation 
functions $\hat{O}$, the ones which can be written as a bulk average of local 
operators $\hat{O}_i$: $\hat{O}= \sum_i \hat{O}_i/L$.
This class of operators includes for instance the average
kinetic or potential energy or the spin-spin correlation function 
at a given distance $d$, namely $\hat{O}_i= \hat{{\bf S}}_i \cdot 
\hat{{\bf S}}_{i+d} $. If we use periodic 
boundary conditions the expectation value of $\hat{O}_i$ 
on a state with given momentum  does not depend on $i$ and the  
bulk average  does not represents an approximation 
\begin{equation}\label{defc} 
{\langle \psi_0 | \hat{O}_i |\psi_0\rangle  \over \langle \psi_0 |\psi_0\rangle  } = 
{\langle \psi_0 | \hat{O} |\psi_0\rangle  \over \langle \psi_0 |\psi_0\rangle  } = C.
\end{equation}
We  show here that  the  expectation value of bulk-averaged operators $\hat{O}$ on the approximate 
state $|\psi_p\rangle$ satisfy the following relation:  
\begin{equation}\label{approx}   
\langle \psi_p |\hat{O}|\psi_p\rangle = C+ O(\epsilon_p^2,\epsilon_p/\sqrt{L}),
\end{equation}
thus implying that for large enough size the  
expectation value (\ref{approx})  approaches  the 
exact correlation function  $C$ linearly with the variance.
The validity of  the above statement is  very simple to show under 
very general grounds.
In fact by definition:
\begin{equation}
\langle \psi_p|\hat{O}|\psi_p\rangle  = C+ 2 \epsilon_p \langle \psi^\prime|\hat{O}|\psi_0\rangle  + \epsilon_p^2 
\langle \psi^\prime|\hat{O} |\psi^\prime\rangle. 
\end{equation}
The term proportional to $\epsilon_p$ in the above equation can be 
easily bounded by use of the Schwartz inequality:
\begin{equation}
|\langle \psi^\prime|\hat{O}|\psi_0\rangle |^2 = |\langle \psi^\prime|\hat{O}-C |\psi_0\rangle |^2 \le 
\langle \psi_0|(\hat{O}-C)^2 |\psi_0\rangle~.  
\end{equation}
The final term in the latter inequality can be estimated  
under the general assumption that correlation functions
$C(d)= \langle \psi_0 |(\hat{O}_i-C)  (\hat{O}_{i+d}-C) |\psi_0\rangle / \langle\psi_0|\psi_0\rangle $ 
decay sufficiently fast 
with the distance $|d|$, as a consequence of the cluster property:
$$\langle \psi_0|(\hat{O}-C)^2 |\psi_0\rangle  = (1+\epsilon_p^2)  {1\over L} \sum_d C(d).$$
This concludes the proof of the statement of this Appendix, provided 
$\sum_d C(d)$ is finite for $L\to \infty$.

\begin{acknowledgments}
This work was partially supported by MURST (COFIN99) and INFM.
Useful discussions with T.M. Rice, M. Troyer, C. di Castro, 
A. Parola, M. Calandra, and M. Capone are gratefully acknowledged.
\end{acknowledgments}

\end{document}